\documentclass[useAMS,usenatbib,onecolumn]{mn2e}
\usepackage{epsfig,amssymb,amsmath,graphicx,enumitem}

\newcommand{\be}{\begin{equation}}
\newcommand{\ee}{\end{equation}}

\newcommand{\rstar}{R_*}
\newcommand{\mstar}{M_*}

\title[GW upper limits on internal magnetic field strength of pulsars]{Updated gravitational-wave upper limits on the internal magnetic field strength of recycled pulsars}
\author[A. Mastrano and A. Melatos]{A. Mastrano$^{1}$\thanks{E-mail:
alpham@unimelb.edu.au} and A.
Melatos$^{1}$\thanks{E-mail: amelatos@unimelb.edu.au}\\
$^{1}$School of Physics, University of Melbourne, Parkville VIC
3010, Australia}

\begin{document}

\date{Accepted ?. Received ?; in original form ?}

\pagerange{\pageref{firstpage}--\pageref{lastpage}} \pubyear{?}

\maketitle

\label{firstpage}

\begin{abstract}

\noindent{Recent calculations of the hydromagnetic deformation of a stratified, non-barotropic neutron star are generalized to describe objects with superconducting interiors, whose magnetic permeability $\mu$ is much smaller than the vacuum value $\mu_0$. It is found that the star remains oblate if the poloidal magnetic field energy is $\gtrsim 40\%$ of total magnetic field energy, that the toroidal field is confined to a torus which shrinks as $\mu$ decreases, and that the deformation is much larger (by a factor $\sim \mu_0/\mu$) than in a non-superconducting object. The results are applied to the latest direct and indirect upper limits on gravitational-wave emission from Laser Interferometer Gravitational Wave Observatory (LIGO) and radio pulse timing (spin-down) observations of 81 millisecond pulsars, to show how one can use these observations to infer the internal field strength. It is found that the indirect spin-down limits already imply astrophysically interesting constraints on the poloidal-toroidal field ratio and diamagnetic shielding factor (by which accretion reduces the observable external magnetic field, e.g. by burial). These constraints will improve following gravitational-wave detections, with implications for accretion-driven magnetic field evolution in recycled pulsars and the hydromagnetic stability of these objects' interiors.}

\end{abstract}


\section{Introduction}

The external magnetic field of a neutron star is (relatively) easily inferred from its spin-down rate, but its internal magnetic field is not directly observable. The main clue suggesting the existence of strong internal neutron star fields comes from the 1998 August 27 giant flare from the soft gamma-ray repeater (SGR) 1900+14 \citep{fetal99,hetal99,metal99}. The giant flare, which released $\sim\ 10^{37}$ J of energy as X-rays, was accompanied by a 2.3-fold increase in the spin-down rate \citep{metal99,wetal99,tetal00}. To explain this, \citet{i01} proposed that the flare and the enhanced spin down were caused by a global reconfiguration of the internal magnetic field of $\sim 10^{13}$ T, well above the external dipole field of $6.4\times 10^{10}$ T\footnote{Throughout this paper, we use SI units: 1 T $= 10^{4}$ G.}. \citet{co11} generalised the \citet{i01} calculation (by allowing the toroidal field strength to change, as well as the moment of inertia) and concluded that an internal field strength of $\sim 10^{12}$ T gave rise to the 1998 August 27 event, lower than the first estimate by \citet{i01}, but still significantly higher than the observed external field.

Stellar ellipticity can also be used to constrain the strength of a star's internal field \citep{c02,detal09,aetal10,p11}. It is well known that a strong magnetic field can deform a star \citep{cf53,f54,g72,k89,pm04,hetal08,metal11}. The ellipticity $\epsilon$ is roughly proportional to the magnetic energy \citep{c02,hetal08,detal09}. Neutron stars, with their intense magnetic fields, possess significant ellipticities, making them good candidates for gravitational wave sources \citep{bg96,mp05,setal05,hetal08,detal09}. Recent data from the fifth Laser Interferometer Gravitational Wave Observatory (LIGO) Science Run set an upper limit of $\epsilon\lesssim 1.4\times 10^{-4}$ on the Crab Pulsar \citep{aetal08,aetal10}, translating into an internal magnetic field of $\lesssim 10^{12}$ T under standard assumptions. LIGO non-detections of the central compact object (CCO) in the supernova remnant Cassiopeia A (Cas A) have constrained its ellipticity as well. The Cas A CCO has not been detected electromagnetically, making it impossible to infer its external magnetic field from the spin-down rate. However, \citet{wetal08} and \citet{w10} constrained its ellipticity as a function of gravitational wave frequency (e.g., $\epsilon\lesssim 3.6 \times 10^{-4}$ for 100 Hz, $\epsilon\lesssim 0.6\times 10^{-4}$ for 200 Hz, and $\epsilon\lesssim 0.38\times 10^{-4}$ for 300 Hz), implying an internal magnetic field $\lesssim 10^{14}$ T. Lastly, \citet{cetal11} showed that it will be possible to use future data from LIGO to set a lower limit of $10^7$ T on the magnetic field ($\epsilon \lesssim 10^{-4}$) of the putative 24-year-old neutron star in the supernova remnant SNR 1987A.

Gravitational waves are generated by a rotating star when it is not spherically symmetric and when its `wobble angle' $\theta$, the angle between its total angular momentum vector and symmetry axis, is nonzero. In general, therefore, the star precesses as it radiates. The magnetic field analysed in this paper is axisymmetric, which deforms the star into an ellipsoid. The most general expression for the gravitational wave signal of a precessing triaxial ellipsoid [given by, e.g., Eqs. (19)--(26) of \citep{jks98}] depends on $\theta$ and the angle $\iota$ between the angular momentum vector and line of sight to the Earth, as well as $\epsilon$. The signal is strongest when $\theta=\pi/2$ (i.e., the rotation axis is perpendicular to the symmetry axis, which is in turn parallel to the magnetic axis) and $\iota=0$ (i.e., the rotation axis is directed towards Earth). In this paper, to simplify matters, we henceforth assume implicitly that $\theta=\pi/2$ and $\iota=0$, so that gravitational wave emission and signal detection are assumed to be optimal and there is no precession, in order to focus on $\epsilon$ without geometric complications.

In \citet{metal11}, we constructed hydromagnetic equilibria for stratified, \emph{non-barotropic} stars. The commonly adopted barotropic assumption, while simplifying calculations, severely restricts the form of the field that can be `fitted' into the star, e.g., \citet{hetal08} found that the field must vanish at the surface, contrary to observations, and \citet{lj09} and \citet{cfg10} found that only configurations dominated by the poloidal component (poloidal energy $\gtrsim 90\%$ of total) are allowed, contrary to the numerical simulations of \citet{bn06}. By abandoning the barotropic assumption, we are able to construct a simple, self-consistent hydromagnetic equilibrium with an internal field that can be matched to an external dipole. In our configuration, because we do not require the pressure to be a function of density alone, we are less restricted in the choices of poloidal and toroidal components; they need not be of any particular relative strengths and are independently adjustable. Incidentally, this means that the magnetic field configuration is independent of the equation of state chosen, cf. \citet{lj09} and \citet{cfg10}.

In this paper, we present one possible astrophysical application of the aforementioned result, namely to constrain the internal magnetic fields of neutron stars in conjunction with gravitational wave observations. As we can match the internal field to an external dipole, we are able to relate one set of observations (external magnetic field strength, from spin period and spin-down rate) with another (gravitational wave upper limits), at least in principle. In Sec. 2, we briefly describe our field structure and summarize the results of \citet{metal11}. Then, in Sec. 3, we generalize the earlier work to include a superconducting interior. We compare the ellipticity calculated using our superconducting model to the gravitational-wave upper limits of 81 known millisecond pulsars, including also the important effect of accretion-induced diamagnetic shielding. Lastly, in Sec. 4, we summarize our results and discuss the possibility that the internal fields of millisecond pulsars are stronger than currently thought.

\section{Field strength versus ellipticity for non-barotropic stars}

\citet{metal11} considered a general class of poloidal-toroidal magnetic field configurations, which are broadly representative of the field structures observed in numerical simulations \citep{bn06,bs06,b09} and exhibit the following properties:

\begin{itemize}[labelindent=\parindent,leftmargin=*]
\item the field is axially symmetric around the $z$-axis;
\item the poloidal part is continuous with a dipole field outside the star (so there are no surface currents) and vanishes at the circle $(r,\theta)=(0.78R,\pi/2)$ (where $r$ and $\theta$ are the radial and polar coordinates and $R$ is the stellar radius; this locus is called the neutral circle);
\item the toroidal component is confined to the region of closed poloidal field lines around the neutral circle;
\item the current density remains finite and continuous everywhere in the star.
\end{itemize}
We write the magnetic field in the form pioneered by \citet{c56},

\be {\bf{B}}=B_0[\eta_p\nabla\alpha(r,\theta)\times\nabla\phi + \eta_t\beta(\alpha)\nabla\phi],\ee
where $\eta_p$ and $\eta_t$ are dimensionless parameters which define the relative strengths of the poloidal and toroidal components respectively. The function $\beta(\alpha)$ takes the form $\beta(\alpha)=(\alpha-1)^2$ for $\alpha\geqslant 1$ and $\beta(\alpha)=0$ elsewhere, confining the toroidal field to the region where $\alpha$ exceeds unity, the value taken by $\alpha$ at $(r,\theta)=(1,\pi/2)$; the current density goes continuously to zero at this boundary. The flux function $\alpha(r,\theta)$ is taken to be $f(r)\sin^2\theta$. Note that this particular form of $\alpha(r,\theta)$ is only applicable when we try to match our field to an external dipole; other multipoles match different $\theta$-dependences. The radial dependence of $\alpha$ is given by

\be f(r)=\frac{35}{8}\left(r^2-\frac{6r^4}{5}+\frac{3r^6}{7}\right).\ee
The function $f(r)$ is postulated to be of this form to ensure that the field described by Eqs. (1)--(2) is continuous with a dipole field outside the star, that there are no surface currents, and that the current density is finite at the origin [for a more thorough derivation, see \citet{aetal11}].

\begin{figure}
\centerline{\epsfxsize=8cm\epsfbox{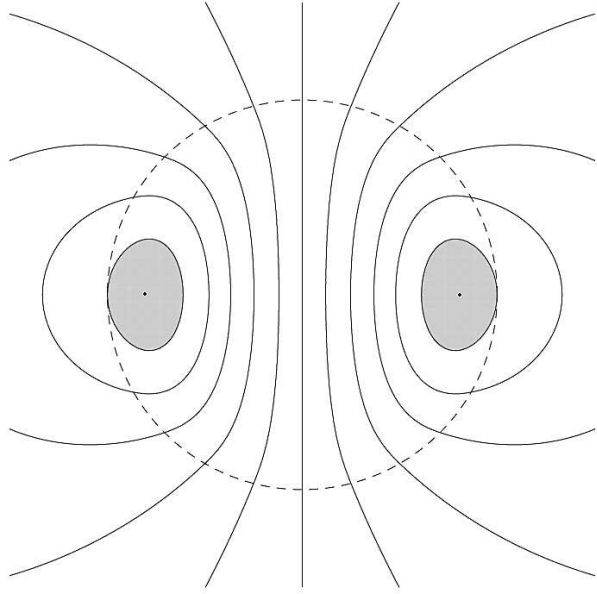}}
 \caption{A cross-sectional diagram of the magnetic field structure analysed in this paper. The surface of the star is represented by the dashed circle. The toroidal magnetic component is confined to the shaded region and fills a torus around the $z$-axis. The poloidal field lines vanish at the neutral circle, i.e. at the dots in the shaded regions.}
 \label{field}
\end{figure}

A schematic diagram of the field is shown in Fig. \ref{field}.\footnote{The stability of this configuration is examined by \citet{aetal11}.} It represents just one possibility amongst many, chosen for simplicity and mathematical convenience; possible configurations involving higher multipoles are not ruled out by observations \citep{a93,tlk02}.

We now calculate the small changes $\delta p$ and $\delta\rho$ to the pressure and density of a star in hydrostatic equilibrium caused by this field, which satisfy the force balance equation \citep{ep77,aw08}

\be (\nabla\times {\bf{H}})\times{\bf{B}}=\nabla\delta p+\delta\rho\nabla\Phi.\ee
In Eq. (3), ${\bf{H}}={\bf{B}}/\mu$ is the magnetic intensity, $\Phi$ is the gravitational potential (the Cowling approximation has been taken, with $\delta\Phi=0$), and $\mu$ is the magnetic permeability. This allows for a superconducting interior (where $\mu$ is smaller than the vacuum permeability $\mu_0$), as analysed in Sec. 3. The assumption that the changes to density and pressure are small enough that they can be treated as perturbations on the steady state is justified \emph{a posteriori}. Ellipticity is then calculated from the perturbed density. Again, the non-barotropic assumption is essential. If barotropy is assumed instead, pressure must be a function of density only, to all orders; a restriction is then imposed on the magnetic field configuration, because the poloidal and toroidal components must be related in such a way that $\delta p$ is purely a function of $\delta\rho$ [e.g., \citet{hetal08}, \citet{lj09}, and \citet{cfg10}].

For the initial unperturbed hydrostatic equilibrium, we adopt a parabolic density profile

\be \rho =\rho_c (1-r^2),\ee
where $\rho_c=15\mstar/(8\pi\rstar^3)$ is the density at the core, $\mstar$ is the mass of the star, and $\rstar$ is the radius of the star. While this is one particular, simple choice of density profile, chosen to render the calculations tractable, we showed in Sec. 3.3 of \citet{metal11} that the resulting ellipticity is within 5\% of that obtained using the more common $n=1$ polytrope.

We define the parameter $\Lambda$ as the ratio of poloidal field energy to total field energy. The energies are obtained by integrating the squares of the poloidal and total magnetic intensities over the star; note that this definition is slightly different from that given by \citet{metal11}, who integrated ${\bf{H}}^2$ over all space. We have $\Lambda=1$ for a purely poloidal field configuration and $\Lambda=0$ for a purely toroidal configuration. In terms of $\Lambda$, for a \emph{non-superconducting star} with $\mu=\mu_0$, the ellipticity $\epsilon$ takes the form

\be \epsilon=5.63\times10^{-6}\left(\frac{B_{\mathrm{s}}}{5\times 10^{10}\textrm{T}}\right)^2\left(\frac{\mstar}{1.4M_\odot}\right)^{-2}\left(\frac{\rstar}{10^4\textrm{m}}\right)^4\left(1-\frac{0.351}{\Lambda}\right),\label{ell}\ee
where $B_{\mathrm{s}}$ is the surface magnetic field strength at the equator, and $M_\odot$ is the Solar mass. As expected, one finds $\epsilon\propto B_\mathrm{s}^2$. The mass quadrupole moment vanishes for $\Lambda=0.351$, that is, when the poloidal field energy is 35.1\% of the total magnetic field energy. Note that both our model and the generalised `twisted torus' model of \citet{cfg10} predict $\epsilon\sim 4\times 10^{-6}$ for a canonical `magnetar-like' neutron star with a purely poloidal field of strength $B_\mathrm{s}=5\times 10^{10}$ T.

\section{Millisecond pulsars}

Armed with Eq. (5), we can use spin-down measurements of $B_{\mathrm{s}}$ and gravitational-wave upper limits on $\epsilon$ to constrain $\Lambda$ and hence the internal magnetic field for various classes of object.

%Now, as another example, if the Crab Pulsar is assumed to have a purely poloidal field structure, with $B_{\mathrm{s}}=3.8\times 10^8$ T (as inferred from pure dipole braking), $\mstar=1.4 M_\odot$, and $\rstar=10^4$ m, Eq. (\ref{ell}) predicts $\epsilon=2.11\times 10^{-10}$.

\begin{figure}
\includegraphics[scale=0.5]{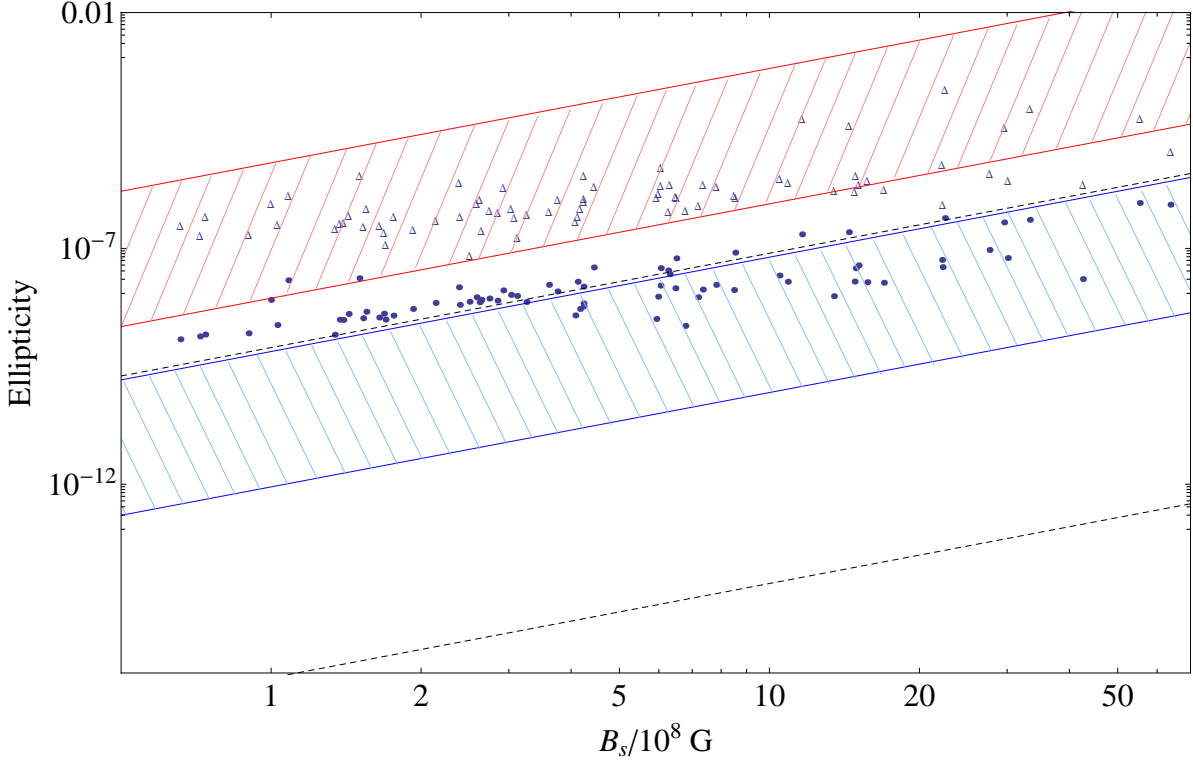}
%\centerline{\epsfxsize=12cm\epsfbox{ligolimitlog.eps}}
%\includegraphics{nscutaway.eps}% Here is how to import EPS art
\caption{Direct LIGO upper limits (triangles) and indirect radio timing spin-down limits (dots) on the absolute ellipticities $|\epsilon|$ of 81 known millisecond pulsars as a function of their measured surface magnetic field strengths $B_\mathrm{s}$ \citep{aetal10}. Also shown are theoretical $|\epsilon(B_\mathrm{s})|$ curves for the following structural and evolutionary scenarios: non-superconducting ($\mu=\mu_0$) (the dashed black curves), and superconducting ($\mu=10^{-3}\mu_0$) with diamagnetic shielding (red and blue curves). The non-superconducting curves are for $\Lambda=10^{-3}$ (realistic; lower dashed black curve) and $\Lambda=10^{-10}$ (unrealistic; upper dashed black curve). The red and blue bands correspond to diamagnetic shielding factors $\xi=B_{\mathrm{s, observed}}/B_{\mathrm{s, actual}}=10^{-4}$ and $10^{-2}$ respectively. Specifically we have $\Lambda=10^{-3}$, $\xi=10^{-4}$ (upper solid red curve), $\Lambda=1$, $\xi=10^{-4}$ (lower solid red curve), $\Lambda=10^{-3}$, $\xi=10^{-2}$ (upper solid blue curve), and $\Lambda=1$, $\xi=10^{-2}$ (lower solid blue curve). Note that all the $\Lambda=1$ curves correspond to oblate stars and all the $\Lambda=10^{-3}$ curves correspond to prolate stars.}
\label{eps}
\end{figure}

In Fig. \ref{eps}, we plot the most up-to-date LIGO (triangles) and spin-down (dots) gravitational-wave upper limits on the absolute value of $\epsilon$ for 81 known millisecond radio pulsars as a function of $B_\mathrm{s}$ \citep{aetal10}. As expected, the indirect spin-down limits are uniformly tighter than the direct LIGO limits for now, although this will change in the future; already, LIGO has beaten the spin-down limits for a handful of non-millisecond pulsars like the Crab \citep{aetal08} and PSR J0537$-$6910 \citep{aetal10}. However, the LIGO limits are included in Fig. \ref{eps} to show, in the discussion below, what constraints can be extracted from gravitational-wave detections when a spin-down measurement is not available, e.g. for a CCO like Cas A. Fig. \ref{eps} also displays a selection of theoretical curves $\epsilon(B_\mathrm{s})$, seen as diagonal lines in the figure. The curves encompass a range of evolutionary scenarios discussed in turn in the subsections below.

Let us begin by considering the lower black dashed curve, which corresponds to Eq. (5) with $\mu=\mu_0$ and $\Lambda=10^{-3}$. In this scenario, the star is not superconducting, the external field is not reduced by accretion in any way, and the internal toroidal field is relatively strong, with the poloidal component contributing only $10^{-3}$ to the total magnetic energy. Even so, $\epsilon$ lies $\sim 6$ orders of magnitude below the strictest of the spin-down limits.

Next, consider the upper black dashed curve, which corresponds to Eq. (5) with $\mu=\mu_0$ and $\Lambda=10^{-10}$. Now the theoretical curve sits comfortably in the region spanned by the spin-down data points. However, this curve represents a physically extreme case, where the toroidal field strength is $\sim 1\%$ of the virial field [$\sim 10^{14}(\mstar/1.4M_\odot)(\rstar/10^4\mathrm{ m})^{-2}$ T \citep{lp07}] and the toroidal field energy is $\sim 10^{10}$ times the poloidal field energy. The value of $\Lambda$ is far below the stable lower limit of $10^{-3}$ calculated by \citet{b09}, suggesting that the upper black dashed curve cannot be realised in practice.

Two important extra effects act in concert to bring the theoretical curves close to the data without appealing to extreme situations like the one in the previous paragraph. First, all objects in Fig. \ref{eps} are intentionally selected to be recycled. It is likely that the actual internal magnetic field is much stronger than the measured external dipole field in a recycled pulsar, because surface currents are diamagnetically shielded \citep{b74}, buried by polar accretion \citep{pm04,mp05}, or resistively dissipated \citep{r90}. Second, it is also likely that the core of the star is a superconductor \citep{j75,ep77,c02,w03,aw08}. We now discuss these scenarios in turn.

\subsection{Core superconductivity}

If the stellar core [$\rho_\mathrm{c}\gtrsim 2.8\times 10^{17}$ kg m$^{-3}$ \citep{lsy99}] is made of a type II superconductor \citep{bpp69,bp75,ep77,eetal96,j06,bs07}, the magnetic permeability $\mu\sim 10^{-3} \mu_0$ drops significantly \citep{ep77,aw08}, and $\delta p$ and $\delta\rho$ are magnified by the factor $\mu_0/\mu$. \citet{aw08} constructed a model of such a star, where the superconducting core is surrounded by a Newtonian fluid, but their model assumes that the star is barotropic and the internal field is purely toroidal. We can generalize their result to our particular non-barotropic configuration. For simplicity, we assume that the entire star is superconducting in this first attempt, instead of just a core region up to $0.6\rstar$ \citep{yls99}.

\begin{figure}
\centerline{\epsfxsize=12cm\epsfbox{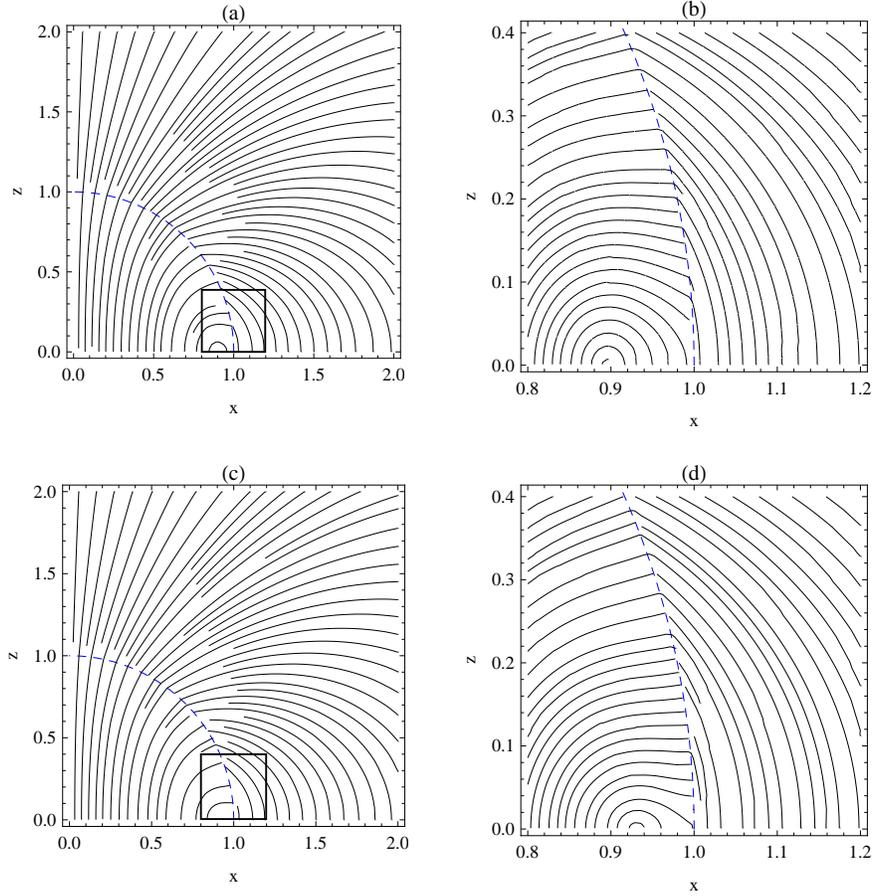}}
 \caption{Poloidal field lines inside and outside a superconducting star with $\mu_r=10^{-1}$ [(a) and (b)] and $\mu_r=10^{-3}$ [(c) and (d)]. The surface of the star is represented by the dashed hemisphere. The right panes (b) and (d) zoom in to the boxed region around the neutral circle in the left panels (a) and (c) respectively. Note that all field lines are, in fact, closed, since the magnetic field is divergence-free.}
 \label{fpt}
\end{figure}

Firstly, to ensure that the stellar field still possesses the properties\footnote{Specifically, we must ensure that the normal component of the internal field and the tangential component of the magnetic intensity are continuous with the external dipole field at $r=1$.} outlined in Sec. 2, the function $f(r)$ in Eq. (2) must be modified into

\be f(r)=\frac{1}{8}[(26+9\mu_r)r^2-(28+14\mu_r)r^4+(10+5\mu_r)r^6],\ee
with $\mu_r=\mu/\mu_0$ [note that this reduces to Eq. (2) for $\mu_r=1$]. The magnetic field is modified in two ways: (1) the field lines become more radial just inside the surface; and (2) the region which contains the toroidal field squeezes closer to the surface and shrinks. In Fig. \ref{fpt}, we sketch the field lines for $\mu_r=10^{-1}$ [Fig. \ref{fpt}(a)] and $10^{-3}$ [Fig. \ref{fpt}(c)]. Recall that the toroidal field is confined to the region where $f(r)\sin^2\theta\geqslant 1$ (this is also the region described by the outermost poloidal field line which closes inside the star). As evident from Eq. (6), this region shrinks as $\mu_r$ decreases (but does not vanish for $\mu_r=0$, the case of perfect superconductivity, tending instead to $\sim 7.5\times 10^{-3}$ of the total volume in the limit $\mu_r\rightarrow 0$). We magnify this region in Fig. \ref{fpt}(b) (for $\mu_r=10^{-1}$) and (d) (for $\mu_r=10^{-3}$). The volume of the torus is 0.16 (0.04) of the original $\mu_r=1$ case for $\mu_r=10^{-1}$ ($\mu_r=10^{-3}$).

The magnetic intensity $\bf{H}={\bf{B}}/\mu$ in Eq. (3) is magnified by $\mu/\mu_0$ in a superconductor. The deformation caused by this field is then calculated by the method given by \citet{metal11}. We are unable to derive a simple analytic formula describing the dependence of $\epsilon$ on $\mu_r=\mu/\mu_0$ and $\Lambda$ simultaneously, but we find that the general form of $\epsilon(\Lambda)$ for a given $\mu_r$ is still similar to Eq. (5), namely

\be \epsilon=c_1\left(\frac{B_{\mathrm{s}}}{5\times 10^{10}\textrm{T}}\right)^2\left(\frac{\mstar}{1.4M_\odot}\right)^{-2}\left(\frac{\rstar}{10^4\textrm{m}}\right)^4\left(1-\frac{c_2}{\Lambda}\right).\label{ellgen}\ee
The dimensionless constants $c_{1,2}$ are quoted in Table 1 for $\mu_r=0.5$, $0.1$, $10^{-2}$, $10^{-3}$, and $10^{-4}$. For all $\mu_r$, the star is oblate for $\Lambda\gtrsim 0.4$ and prolate for $\Lambda\lesssim 0.4$. The functional dependence of $\epsilon$ on $\Lambda$ stays roughly the same as $\mu_r$ changes. The smaller magnetic permeability of the superconducting stellar matter enhances the density perturbation by a factor of $\sim \mu_r^{-1}$, as evident from the force balance equation [Eq. (3)], and this is embodied in the approximate scaling $c_1\propto \mu_r^{-1}$ for small $\mu_r$.

\begin{table*}
 \centering
 \begin{minipage}{170mm}
  \caption{Dimensionless constants $c_{1,2}$ in Eq. (\ref{ellgen}), for some selected values of $\mu_r=\mu/\mu_0$.}
  \begin{tabular}{@{}lcc@{}}
  \hline
     $\mu_r$ &$c_1$& $c_2$\\
\hline
$10^0$ & $5.62\times10^{-6}$ & $0.35$\\
$5\times 10^{-1}$ & $9.97\times10^{-6}$ & $0.42$\\
$10^{-1}$ & $3.92\times10^{-5}$ & $0.40$\\
$10^{-2}$ & $3.84\times10^{-4}$ & $0.42$\\
$10^{-3}$ & $3.83\times10^{-3}$ & $0.43$\\
$10^{-4}$ & $3.84\times10^{-2}$ & $0.43$\\

\hline
\end{tabular}
\end{minipage}
\end{table*}

We can now ask how much closer the theoretical curves approach the data when superconductivity is included. Looking at Fig. \ref{eps} for example, the curve for (say) $\mu_r=10^{-3}$ and $\Lambda=10^{-3}$ (not drawn) is higher than the lower black dashed curve by three orders of magnitude, which is still below observational limits. We conclude that superconducting interiors are easily compatible with current observational upper limits, if the external magnetic field is not reduced by accretion in any way.

We caution that, in this first pass, we assume $\mu_r\neq 1$ throughout the star, instead of only in some region. This assumption is physically implausible and is only taken to simplify our (illustrative) calculation. A more thorough calculation where $\mu_r$ is allowed to vary inside the star is needed before definite conclusions can be drawn regarding the effects of core superconductivity on ellipticity. However, we conjecture that the calculation presented in this section sets the upper limit on the effects of a superconducting interior on stellar deformation: if superconductivity is limited to a smaller region in the star, the changes to $\epsilon$ (relative to $\mu=\mu_0$) will be less than predicted by Eq. (7) and Table 1.

\subsection{Accretion}

Recycled pulsars are selected as the subjects of this study for two reasons: (i) their spin-down rates are lower than those of other objects, yielding more stringent indirect gravitational-wave limits; and (ii) their internal magnetic fields may be much greater than their surface fields due to diamagnetic screening or burial, yielding larger hydromagnetic deformations than one might otherwise expect. We now examine point (ii). Let us ask what happens if the actual magnetic field strength just below the surface takes its pre-accretion value (e.g., before diamagnetic screening or burial) $B_\mathrm{s, actual}=B_\mathrm{s, observed}/\xi$, where $\xi$ is some dimensionless `shielding factor'. We note in passing that this scenario is more realistic than those considered in Sec. 3.1: it is unlikely that the \emph{core} poloidal field is reduced to $\sim 10^4$ T in a recycled pulsar like the surface field, given the high electrical conductivity expected in the core \citep{gr92}, except in the special situation where the source currents reside exclusively in the crust.

We recalculate $\epsilon$ using Eq. (7), substituting $B_\mathrm{s, actual}$ for $B_\mathrm{s}$ and keeping $\mu_r=10^{-3}$. We plot the results for $\xi=10^{-4}$ as the red curves for $\Lambda=1$ (bottom solid red curve) and $\Lambda=10^{-3}$ (top solid red curve) in Fig. \ref{eps}. We also plot $\epsilon$ for the case of $\xi=10^{-2}$ as the blue curves for $\Lambda=1$ (bottom solid blue curve) and $\Lambda=10^{-3}$ (top solid blue curve). For clarity, the case $\mu=\mu_0$ is not presented; it lies three decades lower than the curves with $\mu_r=10^{-3}$. Without a better knowledge of the screening/burial process than is currently at hand, it is best simply to bracket the plausible range $10^{-4}\leqslant \xi\leqslant 10^{-2}$ inferred from population synthesis studies \citep{ketal08}.

Using the potentially stronger shielded pre-accretion fields, the $\epsilon(B_\mathrm{s})$ curves in Fig. \ref{eps} come close to the spin-down limits on $\epsilon$. Now we can see that, aside from perhaps one object, namely PSR J1823--3021A, all the pulsars plotted must have some internal toroidal field component, i.e. $\Lambda<1$, if one has $\xi\leqslant 10^{-2}$. From the red curves and the dots in Fig. \ref{eps}, many objects seem to have $\Lambda<0.01$, even $\Lambda<10^{-3}$ in the notable case of PSR J1910--5959C. On the other hand, the heavily-shielded case $\Lambda=10^{-3}$, $\xi=10^{-4}$ (for example) seems to be ruled out by observations (top solid red curve in Fig. \ref{eps}). In general, $\Lambda$ cannot be too small, for several reasons: it is ruled out by observations, the virial limit sets an absolute upper bound on the internal field strength, and small $\Lambda$ leads to an unstable field configuration \citep{b09}. By contrast, large $\Lambda$ is not ruled out by observations; the only upper bound ($\Lambda=0.8$) is set by the stability analysis of \citet{b09}. Fig. \ref{eps} also tells us that $\xi$ cannot be too small, otherwise the solid red curves exceed the observational upper limits (for $\xi\leqslant 10^{-4}$).

\begin{figure}
\centering
\includegraphics[scale=0.5]{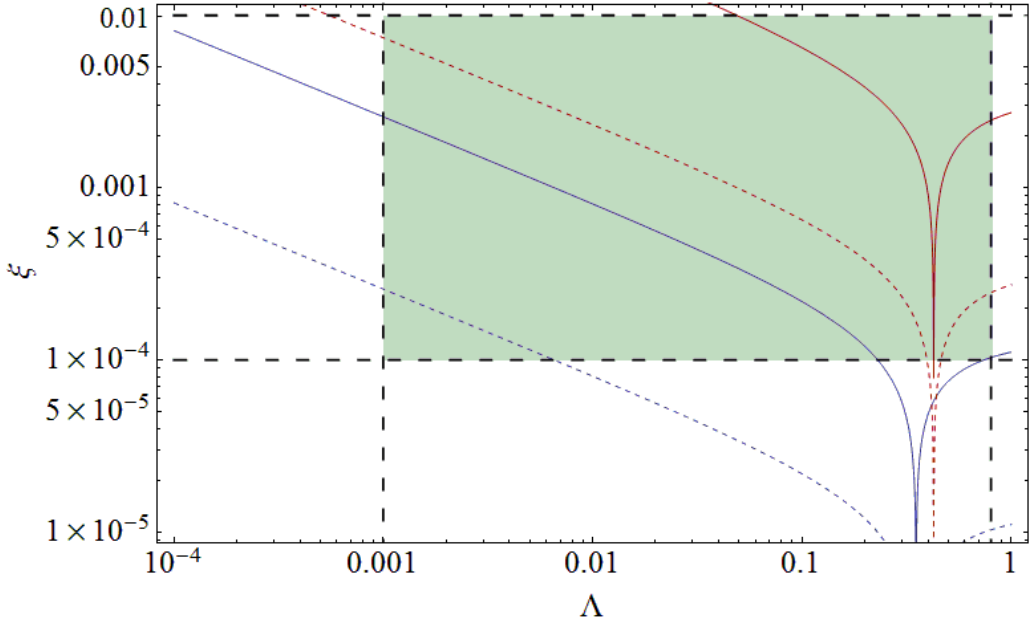}
%\centerline{\epsfxsize=12cm\epsfbox{ligolimitlog.eps}}
%\includegraphics{nscutaway.eps}% Here is how to import EPS art
\caption{Curves of constant ellipticity ($\epsilon=1.184\times 10^{-9}$; strictest limit from Fig. \ref{eps}) on the $\Lambda$--$\xi$ plane, where $\xi$ is a dimensionless `shielding factor', $\xi=B_{\mathrm{s, observed}}/B_{\mathrm{s, actual}}$, and $\Lambda$ is the ratio of the internal poloidal magnetic field energy to the total internal magnetic field energy. Blue curves: non-superconducting $\mu_r=1$, with $B_\mathrm{s, observed} = 10^4$ T (dashed) and $B_\mathrm{s, observed} = 10^5$ T (solid). Red curves: superconducting $\mu_r=10^{-3}$, with $B_\mathrm{s, observed} = 10^4$ T (dashed) and $B_\mathrm{s, observed} = 10^5$ T (solid). The thick black dashed lines are the theoretical bounds on $\Lambda$ and $\xi$: from stability $10^{-3}\leqslant\Lambda\leqslant 0.8$ \citep{b09} and population synthesis $10^{-4}\leqslant\xi\leqslant 10^{-2}$ \citep{ketal08}.}
\label{xip}
\end{figure}

The data in Fig. \ref{eps}, together with the stability-based limits on $\Lambda$ set by \citet{b09}, allow us to use gravitational wave observations and measurements of surface fields to infer bounds on both $\Lambda$ and the shielded field $B_\mathrm{s, actual}$, in principle. We show several possibilities in Fig. \ref{xip}, where we draw curves of constant $\epsilon=1.184\times 10^{-9}$ (corresponding to the current lowest spin-down upper limit, see Fig. \ref{eps}) for a given value of $B_\mathrm{s, observed}$ (solid: $10^5$ T, dashed: $10^4$ T), for $\mu_r=1$ (blue curves) and $\mu_r=10^{-3}$ (red curves). To be consistent with spin-down measurement, an object must lie above the curve relevant to the scenario being considered. The shaded region bordered by thick black dashed lines indicates the theoretical limits on $\Lambda$ [from stability; $10^{-3}<\Lambda< 0.8$ \citep{b09}] and $\xi$ [from population synthesis; $10^{-4}<\xi<10^{-2}$ \citep{ketal08}]. In other words, Fig. \ref{xip} allows us to determine the allowed combinations of $\Lambda$ and $\xi$ for a given observed surface field and $\epsilon$ limit (here $\epsilon=1.184\times 10^{-9}$). For example, a detection from a pulsar with surface field of $10^4$ T indicates $10^{-3}\leqslant\Lambda\leqslant 7\times 10^{-3}$ and $10^{-4}\leqslant\xi\leqslant 2.5\times 10^{-4}$, if we assume the star is non-superconducting (the blue dashed curve), or $10^{-3}\leqslant\Lambda\leqslant 4\times 10^{-1}$ and $10^{-4}\leqslant\xi\leqslant 8\times 10^{-3}$ (prolate star) or $4.5\times 10^{-1}\leqslant\Lambda\leqslant 0.8$ and $10^{-4}\leqslant\xi\leqslant 2.4\times 10^{-4}$ (oblate star) if the star is superconducting with $\mu_r=10^{-3}$ (the red dashed curve). From Eq. (\ref{ellgen}), we see that the oblate cases (with $c_2\leqslant\Lambda\leqslant 1$) are weakly deformed relative to the prolate cases ($10^{-3}\leqslant\Lambda\leqslant c_2$). Hence, for oblate stars to be readily detectable, we need small $\xi$ and $\mu_r$ to boost $\epsilon$ (red curves).

%These data allow us to set the limit $10\lesssim \Lambda (B_\mathrm{s, pre}/B_\mathrm{s})(\mu/\mu_0) \lesssim 10^4$ for recycled pulsars. Note however that this is outside the stable range of $\Lambda$ found by \citet{bn06}, unless their actual field strengths are higher than $10^4$ times their spin-down fields.

%and that our calculated $\epsilon$ are consistent with the LIGO and spin-down upper limits for all 81 objects for reasonable values of $B_\mathrm{s}/B_\mathrm{s, pre}$

\section{Discussion}

This paper makes a first attempt at combining non-barotropic magnetized stellar models \citep{metal11} with LIGO non-detections and radio timing data to constrain the ratio $\Lambda$ and hence the internal magnetic field of recycled pulsars. The models resemble the `twisted torus' of \citet{cetal09}, \citet{lj09}, and \citet{cfg10}, in that they consist of a potentially strong internal toroidal field, which is not observable directly and is confined inside the star, and an external field, which can be measured. The main difference is that we assume the star is stably stratified but non-barotropic \citep{r09}. Because of this, our models allow the poloidal and toroidal components to be adjusted independently of each other; in barotropic models (such as the twisted torus model), the poloidal and toroidal fields must obey certain relations to ensure that pressure is always a function of density purely [see also, e.g., \citet{hetal08}]. This means that our models can easily accommodate field configurations with $0.01\lesssim\Lambda\lesssim 0.8$, the range found to be stable by the numerical simulations of \citet{bn06}.

%If the core fluid of the neutron star is a type II superconductor \citep{bpp69,bp75,ep77,eetal96,j06,bs07}, even larger deformations may result [since, in fact, $\epsilon\propto BH$ and $H\sim 10^3 B$ in a type II superconductor \citep{ep77,aw08}]. \citet{aw08} constructed a model of such a star, where the superconducting core is surrounded by a normal fluid, but their model also assumes the star is barotropic and purely toroidal. However, while a direct comparison between their ellipticity formula [Eq. (195) of \citet{aw08}] and ours [Eq. (5)] is inappropriate, we can infer that application of our model to a superconducting-core neutron star will require a modification of Eq. (5) by a factor of $\lesssim 10^3$.

As seen in Fig. \ref{eps}, our calculated $\epsilon$ is much lower than the LIGO and spin-down upper limits when the inferred dipole field strengths are input directly into Eq. (5).\footnote{Recall that the magnetic axis is assumed to be perpendicular to the rotation axis throughout this paper.} Note that the lower black dashed curve in Fig. \ref{eps} already assumes a strong toroidal field ($\Lambda=10^{-3}$); a weaker toroidal field will generate a curve even lower down. Next, we calculate how interior superconductivity ($\mu_r=10^{-3}$) and accretion-induced screening or burial (red and blue bands in Fig. \ref{eps}) enhance $\epsilon$. This latter possibility is further explored in Fig. \ref{xip}, which shows how, in principle, a gravitational wave detection from a recycled pulsar with a certain surface field (measured from spin down) can be used to infer both $\Lambda$ and shielding factor $\xi$. Fig. \ref{xip} also tells us the sets of parameters that are ruled out by current observations: for a given pulsar with a measured spin-down field strength, non-detection means that $\xi$ and $\Lambda$ lie above that particular curve in Fig. \ref{xip}. For example, the upper solid curve of Fig. \ref{eps}, corresponding to $\xi=10^{-4}$ and $\Lambda=10^{-3}$, lies at the corner of the theoretically allowed shaded region in Fig. \ref{xip} and is therefore ruled out even for a non-superconducting star with a relatively weak observed field strength of $10^4$ T (because it lies below the dashed blue curve in Fig. \ref{xip}).

Is it possible that the magnetic field of a recycled pulsar is much stronger just below the surface than the inferred dipole field? \citet{k91} and \citet{a93}, motivated by the discrepancy between field strengths inferred from dipole spin down and from cyclotron line measurements of accreting X-ray pulsars, raised such a possibility. They proposed that higher-order multipoles may exist close to the surface. \citet{a93} then showed that surface field strengths $\sim 10^4$ times higher than the observed dipole field can account for the aforementioned discrepancy, as well as the location of the millisecond pulsars on the $P$--$\dot{P}$ diagram. The surface field can also be masked by magnetic field burial, whereby accreted matter compresses the polar magnetic flux into a narrow belt around the equator. It has been shown that accretion of $\gtrsim 10^{-5} M_\odot$ is enough to alter the dipole moment significantly \citep{l99,pm04,mp05,zk06,vm09}, depending on the equation of state \citep{pmp11}.

\citet{pg07} raised the possibility that the large-scale magnetic field of a neutron star is supported by long-lived currents in its superconducting core, while small-scale, fast-decaying (decay time $10^5$--$10^7$ yr) magnetic structures also exist near the surface, supported by short-lived currents in the crust. After trying several different initial poloidal and toroidal field strengths, \citet{pg07} found that all their models eventually (after $\sim 1$ Myr) reconfigure into a long-term stable configuration comprising a dipolar poloidal component and a quadrupolar/octupolar toroidal component. The decay of the crustal currents is dominated by Hall drift in the first $10^3$--$10^4$ yr if the initial field is $\gtrsim 10^{10}$ T, i.e. magnetar strength \citep{gr92,rg02,pg07}. When the external dipole field is of magnetar strength, the internal crustal field can still be $\sim 10$ times greater.

\citet{pg07} did not calculate the magnetic deformation of their star, but one can extend their work and that of \citet{metal11} to investigate the effects of this strong, internal, quadrupolar/octupolar toroidal field on stellar ellipticity. To do so, one must remember that Eq. (5) applies to an internal field that is continuous with an external dipole, as shown in Fig. \ref{field}. When the field is distorted significantly by accretion [into, e.g., the `equatorial tutu' shape found by \citet{mp01} and \citet{pm04}], then the internal toroidal field described by Eq. (2) must be recalculated to ensure that the total field still has the desired properties listed at the start of Sec. 2. A similar adjustment must be carried out to accommodate multipolar external fields. It can be shown, for example, that one needs two toroidal field belts around the new neutral circles at $(0.7R,\pi/4)$ and $(0.7R,3\pi/4)$ to match a quadrupolar external field \citep{m10}.

%So far, we seem to accept the existence of a strong internal field without demur, but is it truly possible for a neutron star to possess an internal field much stronger than its observable external field? \citet{k91} and \citet{a93}, motivated by the discrepancy between field strengths inferred from dipole spin down and from cyclotron measurements of accreting X-ray pulsars, raised the possibility that millisecond pulsars have surface fields much stronger than their observed/inferred dipole fields. They proposed that, while the neutron star magnetic field is largely dipolar, higher-order multipoles (which decay with distance much more rapidly than the dipole component) may exist closer to the surface. \citet{a93} then showed that surface field strengths of $\sim 10^4$ higher than the observed dipole field can account for the aforementioned discrepancy, as well as the location of the millisecond pulsars on the $P$--$\dot{P}$ diagram.

Lastly, we remind the reader that a strong internal field is not the only possible cause for stellar deformation. Other effects that can deform a star are:

\begin{itemize}[labelindent=\parindent,leftmargin=*]
\item rotation; ellipticity induced by centrifugal forces is $\epsilon_\Omega\approx 0.3 (\nu/\textrm{kHz})^2$, where $\nu$ is the spin frequency of the neutron star \citep{ps72,c02}. Magnetar spin frequencies are $\sim 0.1$ Hz, giving $\epsilon_\Omega\sim 10^{-9}$. However, the deformation is aligned with the rotation axis [except for a fraction $\sim 10^{-4}$ proportional to the shear modulus \citep{g70}], so it does not contribute to gravitational wave emission or precession \citep{m00};
\item crustal shear stresses; ellipticity supported by the crust's elasticity \citep{hja06}, up to a maximum of $\sim 10^{-7}$ for conventional neutron stars, $\sim 10^{-5}$ for hybrid quark-baryon or meson-condensate stars, or $\sim 10^{-4}$ for solid strange quark stars \citep{o05,hk09}.
\item nuclear processes; a sustained, asymmetric temperature step of $\sim 5\%$ at the base of the crust drives electron capture reactions which produce compositional asymmetries \citep{ucb00}.
\end{itemize}

%$\epsilon_d\approx 6\times 10^{-8}(\nu/\textrm{kHz})^2$ \citep{c02,cul03,hetal08,zbh08,hk09}. For a magnetar, we have $\epsilon_d\sim 10^{-15}$, obviously much smaller than that induced by the magnetic field.

\bsp \label{lastpage}

\end{document}